# Multipass cell for high-power few-cycle compression

Michael Müller,[1,*] Joachim Buldt,[1] Henning Stark,[1] Christian Grebing,[1,2] and Jens Limpert[1,2,3]

[1]*Friedrich Schiller University Jena, Institute of Applied Physics, Albert-Einstein-Straße 15, 07745 Jena, Germany*
[2]*Fraunhofer Institute for Applied Optics and Precision Engineering, Albert-Einstein-Straße 7, 07745 Jena, Germany*
[3]*Helmholtz-Institute Jena, Fröbelstieg 3, 07743 Jena, Germany*
*Corresponding author: michael.mm.mueller@uni-jena.de*

**A multipass cell for nonlinear compression to few-cycle pulse duration is introduced comprising dielectrically enhanced silver mirrors on silicon substrates. Spectral broadening with 388 W output average power and 776 µJ pulse energy is obtained at 82% cell transmission. A high output beam quality ($M^2$<1.2) and a high spatio-spectral homogeneity (97.5%) as well as the compressibility of the output pulses to 6.9 fs duration are demonstrated. Finite element analysis reveals scalability of this cell to 2 kW average output power.**

High-power, few-cycle driving lasers find application, for example, in electron acceleration [1], in extreme-ultraviolet coherent diffractive imaging [2], and in the generation of isolated attosecond pulses [3] used for transient absorption spectroscopy [4]. Most applications can benefit from a higher-power driver to reduce the measurement time and to improve the signal-to-noise ratio.

State-of-the-art few-cycle sources are based on parametric amplification [5,6] or nonlinear compression in noble-gas-filled, hollow capillaries [7–9] driven by ultrafast lasers. Today, for the available kW-level average power, multi-mJ-energy lasers [10–12], both these post-processing techniques become a bottleneck. In particular, the absorption in the nonlinear crystals of parametric amplifiers induces thermal aberrations, clamping their average power to the 100 W-level, and the conversion efficiency from the driving laser to the output signal is low at around 15% [6]. In hollow-core fibers, the transmission is higher ranging from 32% in two-stage [13] to 70% [8] in single-stage systems, which allowed for the demonstration of 318 W average output power [7], but reliable coupling of the driving laser to the hollow-core fiber gets progressively more challenging as the average power increases.

An alternative could be nonlinear compression in multipass cells [14–16], which recently has been demonstrated at 112 mJ pulse energy [17] and at 1 kW average output power [18]. These cells feature an excellent transmission of up to 96% [18] and a high resilience to beam quality imperfections and pointing instabilities due to their extended free aperture, which allows for straightforward power scaling. Hence, these cells were ideal candidates for high-power nonlinear pulse compression to the few-cycle regime, if not the reflection bandwidth of the typical dielectric mirrors used would limit the pulse duration to about 30 fs. In consequence, cascaded cells based on broadband-dielectric [19] and silver-coated mirrors [20] have been investigated, but yielded only mediocre-quality pulses with 18 fs and 13 fs duration, respectively. In Ref. 19, this is due to the oscillations in the group delay dispersion (GDD) of the broadband dielectric mirrors and the large dispersion of the fused silica nonlinear medium. In Ref. 20, it is due to the dispersion of the krypton filling and the large overall compression factor. Furthermore, the low reflectivity of the silver mirrors leads to a low efficiency and the inherent absorption complicates average-power scaling. Thus, it is an open question whether the advantageous high transmission and robustness of multipass cells can be translated to few-cycle pulse compression at all, which is addressed in this contribution.

The following experiment is based on two noble-gas-filled multipass cells of which the second cell features high-power broadband mirrors that are remarkable in two respects: First, the mirrors comprise a dielectrically enhanced silver coating, which minimizes both the reflection loss and the heat load without introducing significant dispersion. Second, the mirrors' substrate material is monocrystalline silicon featuring a low-expansion and an enormous heat conductivity, which allows for effective cooling and which minimizes the absorption-induced thermal aberration. In particular, the laser-heating-induced surface is proportional to the ratio of the substrate's coefficient of linear expansion ($\alpha$) to its heat conductivity ($\kappa$) [21]. Thus, monocrystalline silicon substrates ($\alpha$=2.6 µm/(m·K), $\kappa$=163 W/(m·K) [22]) offer a 25-times weaker thermally-induced aberration compared to fused silica substrates ($\alpha$=0.55 µm/(m·K), $\kappa$=1.38 W/(m·K) [23]).

Figure 1 depicts a schematic of the setup. The driving laser is a coherently combined, ytterbium-based, chirped-pulse fiber amplifier [10]. The laser emits pulses at 500 kHz repetition rate with approximately 1 mJ pulse energy and 200 fs pulse duration. Nonlinear spectral broadening in a first, argon-filled, dielectric-mirror Herriott cell and chirped-mirror compression reduces the pulse duration to 31 fs [18]. After this stage, the collimated beam features a $1/e^2$-diameter of 8 mm and an average power of 474 W. The beam continues into a second multipass cell comprising 13 individual 1"-diameter, dielectrically enhanced silver-coated silicon mirrors ($R_M$>98.5%, |GDD|<10fs$^2$, 800-1300 nm). The mirrors have a radius of curvature of 1 m and are set up 1.95 m apart on two array holders that feature water-cooling of each substrate. The cell eigenmode has a beam radius of 230 µm in the focus and of 1.43 mm on the mirror to which the input beam is mode-matched using an off-axis reflecting telescope [24] comprising dielectric mirrors with radii of curvature of -1400 mm (concave) and +1000 mm (convex). In the cell, the beam goes through 13 foci in an atmosphere comprising 0.5 bar of argon and 0.5 bar of helium. The nonlinear spectral broadening occurs in the argon, while the helium only serves to equalize the pressure to one atmosphere eliminating a residual pressure-induced deformation of the vacuum chamber. At the output, the beam is sampled using a series of three fused silica wedges, while the high-power beam is dumped on a thermal power meter. The beam sample leaves the cell through a 1 mm-thick fused silica window and is collimated by a silver-mirror telescope. Then, it is sent over broadband dielectric, chirped-mirror pairs prior to the analysis to demonstrate compressibility.

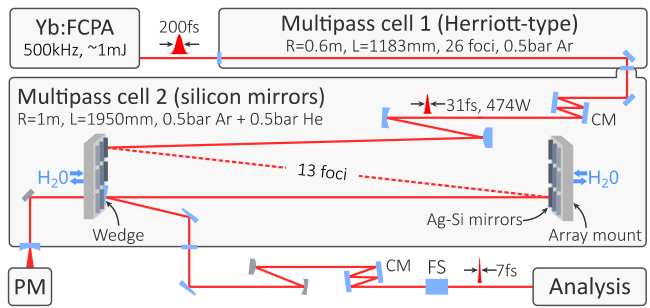

Fig. 1. Schematic of the two-stage multipass cell compression setup. Yb:FCPA: ytterbium-doped fiber chirped-pulse amplifier, R: mirror radius of curvature, L: mirror distance, Ag-Si: dielectrically enhanced silver on silicon, CM: chirped-mirror, PM: power meter, FS: fused silica.

The output of the first compression stage has been characterized in Ref. [18]. The beam features an $M^2$ value of less than 1.15, the pulse is close-to Fourier-limited with a duration of 31 fs, and the effective spatio-spectral homogeneity of the beam is larger than 98.7%. Figure 2 depicts a typical spectrum and a typical autocorrelation trace measured after this first stage in comparison to a simulation based on Ref. [25] that solves the nonlinear envelope equation [26] in radial symmetry [27] including self-phase modulation, self-steepening, diffraction, dispersion to all orders, and the mirror reflection bandwidth. Excellent agreement of measurement and simulation is achieved assuming a Gaussian beam carrying a Gaussian pulse with 994µJ energy, 200 fs duration, and an additional third-order dispersion of -10$^4$ fs$^3$ to reproduce the slight asymmetry in the broadening.

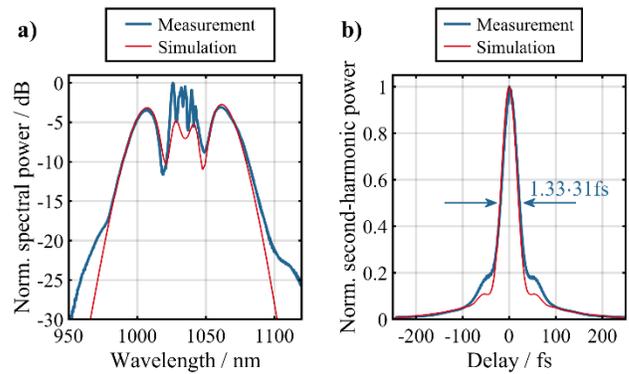

Fig. 2. Measured (blue) and simulated (red) output spectra (a) and autocorrelation traces (b) after the first multipass cell.

The beam continues into the second multipass cell resulting in further spectral broadening. Figure 3 depicts the measured spectrum after the cell. Spectral content down to the -20dB level is generated in the span from 750 nm to 1300 nm. A simulation of this cell, starting with the output of the previous stage, shown in red, yields reasonable agreement. Corrected for the wedge, the output power is 388 W, which corresponds to 82% transmission of the second cell and meets the mirror specification ($R_M^{13}$). This is the same transmission achieved in the broadband-dielectric-mirror cell of Ref. [19] and almost twice the transmission of the protected-silver-mirror cell of Ref [20].

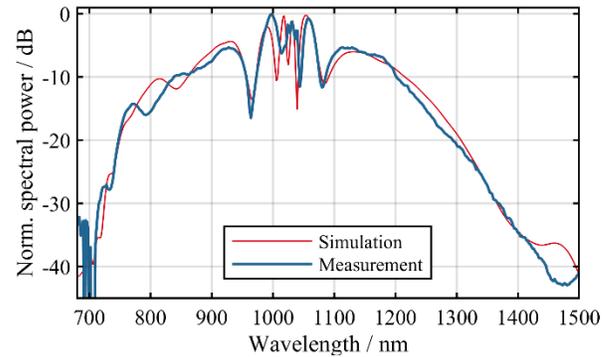

Fig. 3. Measured (blue) and simulated (red) spectra at the output of the second multipass cell.

A dispersion scan [28] of the beam sample is shown in Fig. 4 after compression with -540 fs$^2$ GDD after several reflections on a chirped mirror pair and after transmission through 21 mm fused silica yielding an experimental minimum of second- and third-order dispersion. The retrieval with a root-mean-square error of 0.6% results in a pulse duration of 6.9 fs and 61% relative peak power with respect to the 5.6 fs-transform limit of the spectrum. Linear scaling from this result infers 60 GW peak power for compression of the high-power beam.

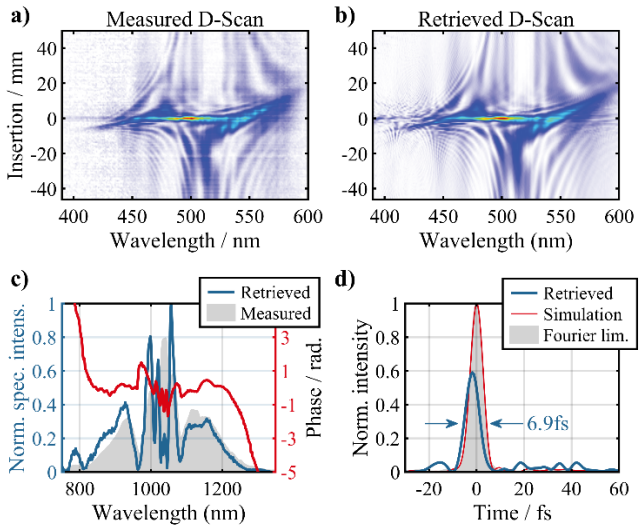

Fig. 4. Measured (a) and retrieved (b) dispersion scan in a false color linear scale, the measured and retrieved fundamental spectra (c) including the retrieved spectral phase, and the reconstructed pulse (d) with 6.9 fs FWHM pulse duration in comparison to the Fourier limit and to the simulation result after ideal compression with negative GDD.

The output beam quality is analyzed by measuring the beam caustic with a silicon-CCD as shown in Fig. 5. Weighting the measured spectrum (Fig. 3) with the spectral response of the CCD reveals that the detected center wavelength is 897 nm to which the measured caustic yields $M^2$ values less than 1.2 for both axes. This result implies that no significant beam quality degradation occurred relative to the input beam [18] at least within the spectral sensitivity range of the CCD (< 1.1 µm). This result is supported by the radially resolved simulation of the nonlinear propagation predicting a negligible $M^2$ degradation.

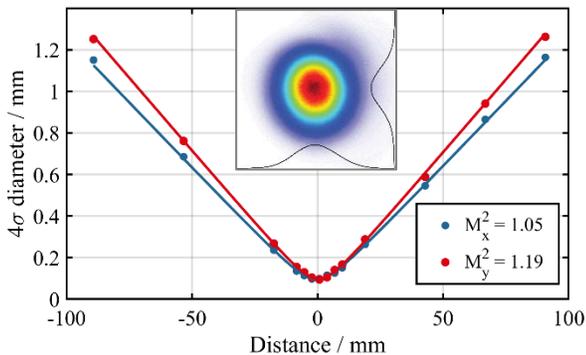

Fig. 5. ISO-conform $M^2$ measurement for the CCD-sensitivity-weighted wavelength of 897 nm. Inset: Collimated beam profile with lineouts.

The spectral homogeneity of the output beam is measured using a 1D-imaging spectrograph comprising a grating with 300 lines per millimeter, a cylindrical lens with 50 mm focal length, and a Si-CCD. Stitching frames with increasing exposure time and correcting for the CCD spectral sensitivity allowed extending the detection range beyond 1.1 µm as shown in Fig. 6. Evaluating the overlap of the off-axis spectra with the on-axis spectrum as defined in Ref. [29] yields an effective homogeneity of 97.5%, proving that no significant degradation occurred with respect to the input [18]. This measurement is a lower-bound estimate as the result is limited by the low signal-to-noise-ratio for wavelengths larger than 1.1 µm.

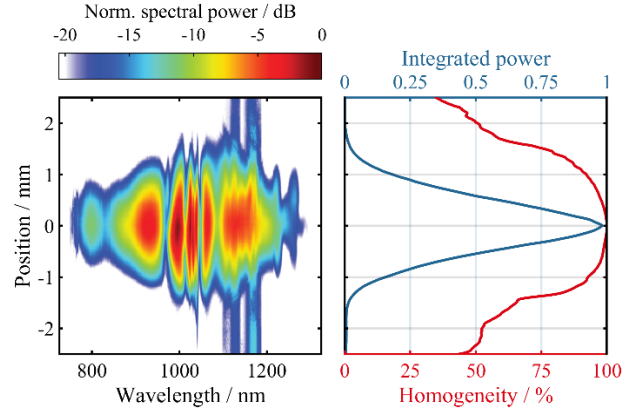

Fig. 6. Measured, CCD-sensitivity-corrected spatio-spectral map (left panel) with a beam lineout and homogeneity (right panel) revealing an effective homogeneity of 97.5%.

Lastly, the power-scaling limit of the second cell imposed by thermal aberration of its mirrors is analyzed using the radially symmetric beam propagation simulation. For that, the thirteen mirrors' internal temperature distribution due to their 1.5% coating absorption and their resulting surface deformation are calculated by finite element analysis (MATLAB 2020b PDE toolbox). Each mirror assembly is represented by a 1"-diameter segmented cylinder as shown in Fig. 7(a), of which the 6.35 mm thick top segment is the silicon substrate, followed by a 200 µm thick layer of glue and the 10 mm thick, water-cooled, aluminum front plate of the mirror mount. From top to bottom, the segments are assigned their heat conductivity (163, 1.5, 120 W/(m·K)), coefficient of linear expansion (2.6, 50, 24 µm/(m·K)), Poisson ratio (0.3), and Young's modulus (113, 4, 70 GN/m$^2$). Two boundary conditions are set for the calculation of the temperature distribution, namely the incoming heat flux in shape of the beam profile and a fixed temperature of the bottom surface (300 K) emulating the contact to the cooling water. An example simulation outcome is depicted in Fig. 7(a) showing the surface temperature for the assembly at an incident power of 2.5 kW. Subsequently, the mirror surface deformation is computed assuming that only the bottom rim is fixed, allowing for free expansion in the mirror normal direction. The resulting wavefront aberration is imposed on the propagating beam. Figure 7(b) displays the highest peak temperature rise of the mirrors and the beam's $M^2$ value after the thirteen focal passes of the cell for increasing input average power. Accepting an $M^2$ degradation of maximal 0.2, which corresponds to 31% reduction of the focusable peak intensity, allows for an input average power of 2.7 kW resulting in 2.2 kW average power output. The water-cooling clamps the peak and mean temperature rise to 80 K and 30 K, avoiding mirror damage and minimizing the convection in the cell, respectively. This simulation underpins the high resilience of monocrystalline silicon to thermal aberration. Hence, these substrates are viable also for the chirped-mirror compressor of the high power beam and for steering mirrors transporting the beam to the experiments thereafter.

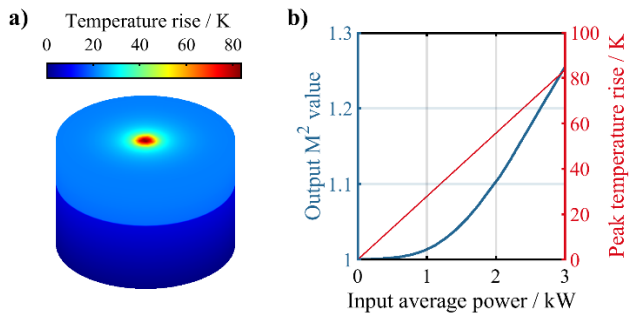

Fig. 7. (a) Model geometry of the silicon mirror glued to a water-cooled aluminum heat sink with a false-color overlay of the surface temperature rise for 3 kW incident average power. (b) Simulation output $M^2$ value (blue) and mirror peak temperature rise (red) for the few-cycle multipass cell at increasing input average power.

In summary, a multipass cell for nonlinear pulse compression to few-cycle pulse duration is demonstrated. The key component are reflectivity-enhanced silver mirrors on water-cooled silicon substrates making excellent high-power broadband reflectors. The achieved output average power of 388 W is a record for broadening schemes supporting few-cycle pulse duration and the compressibility of the 776 µJ-energy output pulses to 6.9 fs duration is demonstrated. The output beam shows neither thermal aberration nor significant spatio-spectral inhomogeneity and finite element analysis indicates aberration-free scalability to about 2 kW average output power.

A particular highlight is the 82% transmission of the cell that, upon inclusion of the 96% transmission of the first compression stage [18], infers a 78% overall transmission from the driving laser to the spectrally broadened output, surpassing even the record transmission in hollow-core fiber compression [8]. Furthermore, the extended free aperture of the cells greatly simplifies power scaling, enabling few-cycle driving lasers of unprecedented average power, efficiency, and reliability.

**Funding.** Fraunhofer-Gesellschaft (Cluster of Excellence "Advanced Photon Sources"); European Research Council (835306).

**Disclosures.** The authors declare no conflicts of interest.